\documentclass[twocolumn,showpacs,preprintnumbers,superscriptaddress]{revtex4}
\usepackage{graphicx}
\usepackage{dcolumn}
\usepackage{bm}
\begin{document}

\def\btt#1{{\tt$\backslash$#1}}
\def\BibTeX{\rm B{\sc In}\TeX}

\title{Resonance instability  of axially-symmetric magnetostatic equilibria} 
\author{Alfio Bonanno}
\affiliation{ INAF, Osservatorio Astrofisico di Catania,
           Via S.Sofia 78, 95123 Catania, Italy} 
\affiliation{INFN, Sezione di Catania, Via S.Sofia 72,
           95123 Catania, Italy} 
\author{Vadim Urpin}
\affiliation{ INAF, Osservatorio Astrofisico di Catania,
           Via S.Sofia 78, 95123 Catania, Italy} 
\affiliation{A.F.Ioffe Institute of Physics and Technology, 
           194021 St. Petersburg, Russia}

\date{\today}
\begin{abstract}
We review the evidence for and against the possibility that a strong enough 
poloidal field stabilizes an axisymmetric magnetostatic field configuration. 
We show that there does exist a class of 
resonant MHD waves which produce instability
for any value of the ratio of poloidal and toroidal 
field strength. 
We argue that recent investigations of the stability of mixed poloidal
and toroidal field configurations based on 3-d numerical simulations, can miss this
instability because of the very large azimuthal wave numbers involved and its resonant character. 
\end{abstract}
\maketitle

\section{Introduction}
The stability of hydromagnetic configurations is still a topic of debate. 
Even simple magnetic configurations consisting in a pure azimuthal (toroidal) 
or vertical (poloidal) field are generally unstable (see, e.g., \cite{frei}), 
yet the magnetic fields  
observed in several astrophysical contexts are stable on a secular time scale.
{ 
In this context, the energy principle of Bernstein et al. \cite{bernstein}  
has extensively been used in the past to study the stability of simple 
poloidal or toroidal fields \cite{wr73,tay73a,tay73b} and also of mixed 
combinations of the two \cite{tay80}. In cylindrical geometry, it can be 
proved that the plasma is stable for all azimuthal and vertical wave numbers 
$m$ and $k$, if it is stable for $m=0$ in the $k\rightarrow 0$ limit,
and for $m=1$ for all $k$ \cite{gopo}. 
On the other hand, to show that a generic configuration with a 
combination of  vertical field and non-homogenous azimuthal field 
is stable against the $m=1$ mode (for all $k$) is not an easy task in general and 
one has to resort  either to a variational approach or to a numerical investigation 
of the full eigenvalue problem in the complex plane \cite{deblank}.
In this respect, the ``normal mode" approach can be more useful in astrophysics, as 
it is often important to know the growth rate of the instability and the properties of the 
spectrum of the unstable modes \cite{bo08a,bo08b}. 
}

In recent years, the use of 3D numerical simulations has opened up the possibility 
of studying the stability of various field configurations following the evolution
from the linear phase to the non-linear regime. A strategy often used is to evolve a generic initial state which 
eventually relaxes to a final configuration assumed to be stable \cite{bs04,bn06,br08,br09,duez}.
The drawback with this  approach is that it is difficult to characterize the 
topology of the final configuration from the analysis of the numerical data and to determine 
a class of sufficient conditions for instability which could be of astrophysical interest. 
In particular, the conclusions of some recent works in this direction seem to 
point out that it is the strength of the poloidal field which stabilizes the basic 
state \cite{br08,br09}. 

The aim of this paper is to clarify that field configurations containing generic combinations
of axial and azimuthal fields are subject to a class of resonant MHD 
waves which can never be stabilized for any value of the 
ratio of poloidal and toroidal fields.  The instability of these waves has a mixed 
character, being both current- and pressure-driven \cite{bo11}. We argue that in 
this case the most dangerous unstable modes are resonant, i.e. the wave vector 
$\vec{k} = (m/s) \vec{e}_\theta + k_z \vec{e}_z$ is perpendicular to the magnetic 
field, $\vec{B} \cdot \vec{k} = 0$ where $k_z$ is the wavevector in the axial 
direction, $m$ is the azimuthal wavenumber, and $s$ is  the cylindrical radius.
The length scale of this instability depends on the ratio of poloidal and 
azimuthal field components and it can be very short, while the width of the 
resonance turns out to be extremely narrow. For this reason its excitation in  
simulations  can be problematic.

{
The paper is organized as follows. In Sec.2, the main equations governing 
the behaviour of linear perturbations in cylindrical plasma configurations are 
presented. In Sec.3, we consider a linear stability analysis of such configurations,
using an analytical approach complemented by a numerical investigations. 
Direct numerical simulations of the non-linear evolution of a
cylindrical configuration are presented in Sec.4. In Sec.5, we compare our
results with those obtained by other authors and discuss possible astrophysical
applications of this instability.}  
\section{Basic equations}
Let us consider an axially symmetric basic state with azimuthal and axial magnetic 
fields. The azimuthal field is assumed to be dependent on $s$ alone, $B_{\varphi}= 
B_{\varphi}(s)$, but the axial magnetic field $B_z$ is constant. We assume that 
the sound speed is significantly greater than the Alfv\'en velocity in order to justify the use of 
incompressible MHD equations 
\begin{eqnarray}
\frac{\partial \vec{v}}{\partial t} + (\vec{v} \cdot \nabla) \vec{v} = 
- \frac{\nabla P}{\rho} 
+ \frac{(\nabla \times \vec{B}) \times \vec{B}}{4 \pi \rho}, 
\nonumber \\
\frac{\partial \vec{B}}{\partial t} - \nabla \times (\vec{v} \times \vec{B}) 
= 0 \;,
\;\;\; \nabla \cdot \vec{B} = 0 \:,\;\;\; \nabla \cdot \vec{v} = 0 .  
\end{eqnarray}
In the basic state, hydrostatic equilibrium in the radial direction is assumed. 
We study a linear stability with respect to small disturbances. Since the basic 
state is stationary and axisymmetric, the dependence of disturbances on $t$, 
$\varphi$, and $z$ can be taken in the form $\exp{(\sigma t - i k_z z - 
i m \varphi)}$. Linearizing Eq.(1) and eliminating all variables in favor of the 
radial velocity disturbance, $v_{1s}$, we obtain 
\begin{eqnarray}
\frac{d}{ds} \left[ \frac{1}{\lambda} (\sigma^2 + \omega_A^2) \left( 
\frac{d v_{1s}}{ds} + \frac{v_{1s}}{s} \right) \right] 
- k_z^2 (\sigma^2 + \omega_A^2) v_{1s} +  
\nonumber \\
2 \omega_{B} \left[ \frac{m (1 + \lambda)}{s^2
\lambda^2} \left( 1 - \frac{\alpha \lambda}{1 + \lambda} \right) (\omega_{Az} +
2 m \omega_{B} ) 
+ \frac{m \omega_{Az}}{s^2 \lambda^2} 
\right.
\nonumber \\
\left. 
- k_z^2 \omega_{B} (1 - \alpha)  \right] v_{1s} + 
\frac{4 k_z^2 \omega_{A}^2 \omega_{B}^2}{\lambda (\sigma^2 + \omega_{A}^2)} v_{1s}
=0, \;\;\;\;\;\;
\label{due}
\end{eqnarray} 
where $\omega_{A}=( \vec{B} \cdot \vec{k} )/
\sqrt{4 \pi \rho}$, $\omega_{Az} =  k_z B_z/ \sqrt{4 \pi \rho}$,
$\omega_{B} = B_{\varphi}/s \sqrt{4 \pi \rho}$,
$\alpha = \partial \ln B / \partial \ln s$, and $\lambda = 1 + m^2/s^2 k_z^2$.
{
Eq.(\ref{due}) describes the stability problem as a nonlinear eigenvalue problem.
This equation has been first derived by Freidberg \cite{frei70} in his study of 
MHD stability of a diffuse screw pinch (see also \cite{bo08b}). The author found
that, for a given value of $k_z$, it is possible to obtain multiple values of the
eigenvalue $\sigma$, each one corresponding to a different eigenfunction, and 
calculated $\sigma$ for the fastest growing fundamental mode. The most general
form of Eq.(2), taking into account compressibility of plasma, was derived by 
Goedbloed \cite{goed71}. Since we study the  stability assuming that the magnetic energy
is smaller than the thermal one, the incompressible form of Eq.(\ref{due}) can be a sufficiently  
accurate approximation.
}
In fact, Eq.(\ref{due}) was studied by Bonanno \& Urpin \cite{bo08b} in their analysis
of the non-axisymmetric stability of stellar magnetic fields. 
%

We can represent the azimuthal magnetic field as $B_{\varphi} = B_{\varphi 0} 
\psi(s)$, where $B_{\varphi 0}$ is its characteristic strength and $\psi (s) 
\sim 1$.  It is convenient to introduce dimensionless coordinate $x= s/s_2$ 
and dimensionless quantities $q = k_z s_2$, $\Gamma  = \sigma/\omega_{B0}$,
$\omega_{B0} = B_{\varphi 0}/s_2 \sqrt{4 \pi \rho}$, and $\varepsilon = B_z/
B_{\varphi 0}$.
Then, Eq.~(2) transforms into
\begin{eqnarray} \label{pert}
\frac{d}{dx} \! \left( \! \frac{d v_{1s}}{dx} \! + \! \frac{v_{1s}}{x} \! \right) \! + \!
\left( \! \frac{d v_{1s}}{dx} \! + \! \frac{v_{1s}}{x} \! \right) \! \frac{d \ln \Delta}{d x} -
\frac{2 q^2 \psi(x)}{x (\Gamma^2 + f^2)} \! \times \!
\nonumber \\
\left\{ \left[ \left( 1 \! - \!
\frac{m^2}{q^2 x^2} \right) \frac{\psi(x)}{x} \!-\!\frac{m \varepsilon}{q x^2} 
\right]
(1 - \alpha) \! - \! \frac{2 m f}{m^2 + q^2 x^2} \right\} v_{1s} 
\nonumber \\
- \! q^2 \left(1 + \frac{m^2}{q^2 x^2} \right) v_{1s} 
+ \frac{4 q^2 f^2 \psi^2(x)}{x^2 (\Gamma^2 + f^2)^2} v_{1s} = 0, \;\;\;\;\;
\end{eqnarray}
where
\begin{equation}
f = q \varepsilon + m \frac{\psi(x)}{x}, \;\; \Delta = 
\frac{q^2 x^2 (\Gamma^2 +
f^2)}{m^2 + q^2 x^2}.  
\end{equation}

{
With appropriate  boundary conditions, Eq.~(3) allows to determine the eigenvalue $\Gamma$. 
If the inner boundary is extended to include the cylinder axis
it is not difficult to show that the eigenfunction for $m=1$ must be non-vanishing there  to ensure regularity.
This result follows from the series solution of Eq.(3) near $x=0$, so that 
$v_{1s} \propto x^{b}$ with  $b= -1 \pm m$, and regularity at $x=0$ implies $b=0$ for $m=1$,
and $b>0$ for $m>1$. In the setup discussed in this paper the inner boundary is not located at the axis, and 
we can safely assume that $v_{1s}=0$ at $x=x_1$  and $x=x_2$.} 
We will demonstrate the occurrence of a resonance instability in magnetic
configurations  by an analytical and numerical solution of 
Eq.~(3), and by 3D direct numerical simulations. 
\section{Linear analysis of instability}
\subsection{Analytical considerations}
{
It is interesting to have a qualitative understanding of the MHD spectrum, thus
solving Eq.~(3) in the small gap approximation.} In this case one assumes that the distance between
the boundaries, $\Delta x=x_2-x_1$, is small compared to 
$x_2=1$ and neglect in Eq.~(3) terms of the order $v_{1s}/x$ compared 
to $d v_{1s}/ dx$. In this approximation, all coefficients of Eq.~(3) can be 
considered as constant and Eq.~(3) yields 
\begin{eqnarray} 
\frac{d^2 v_{1s}}{dx^2} - \frac{2 q^2}{(\Gamma^2 + f^2)} \left[ \left( 1 \! - \!
\frac{m f}{q^2} \right) (1 - \alpha) \! - 
\! \frac{2 m f}{m^2 + q^2} \right] v_{1s} 
\nonumber \\
- \! (q^2  + m^2) v_{1s} + \frac{4 q^2 f^2}{(\Gamma^2 + f^2)^2} v_{1s} = 0, \;\;\;\;\;
\end{eqnarray}
The solution, satisfying the boundary conditions, is $v_{1s} \propto \sin 
[\pi(x-x_1)/ \Delta x]$. The corresponding dispersion relation is biquadratic and can
be easily solved. The solution is
\begin{eqnarray}
\Gamma^2 = - f^2 - \mu \left[ \left(1 - \frac{mf}{q^2} \right) (1 - \alpha)  - 
\frac{2mf}{m^2+q^2} \right] \pm
\nonumber \\
\left\{ \mu^2 \left[ \left(1 - \frac{mf}{q^2} \right) (1 - \alpha)  - 
\frac{2mf}{m^2+q^2} \right]^2 + 4 \mu f^2 \right\}^{1/2},
\end{eqnarray}
where $\mu = q^2 / [q^2 + m^2 + (\pi/\Delta x)^2]$. The parameter $f$ 
characterizes departures from the magnetic resonance, $\omega_A =0$. 
To show the occurrence of 
instability, we consider solution (6) at small departures from the magnetic 
resonance, $f \approx 0$. If $\alpha > 1$, we have
\begin{equation}
\Gamma^2=\frac{2 m^2 (\alpha-1)}{m^2+(p^2+m^2) \varepsilon^2}
\label{sette}
\end{equation}
where $p^2=(\pi/\Delta x)^2$ and the instability is never suppressed for any finite value of $\varepsilon$.
The growth rate is a rapidly increasing function of $m$
and $\Gamma^2 \approx 
(1+\varepsilon^2)^{-1}$ in the limit $m \gg p^2$.  
If $\alpha < 1$, then Eq.~(6) yields
\begin{equation}\label{ggr}
\Gamma^2 \approx f^2 \frac{1 + \alpha}{1 - \alpha},
\end{equation} 
that implies instability  if $\alpha > -1$. 
The profile with 
$\alpha < -1$ is stable in the small gap limit. Note that modes with $q$
satisfying the resonance condition $\omega_A = 0$ (or $f=0$) are marginally stable 
because $\Gamma=0$ for them, but $\Gamma^2 > 0$ in a neighborhood of the 
resonance. Therefore, the dependence of $\Gamma$ on $q$ should have a two-peak 
structure for any $m$. As in the case $\alpha > 1$, the instability occurs for any value of  
$\varepsilon$. If $\alpha =1$, then we have
\begin{equation}
\Gamma^2 \approx \mu f \left[ \frac{2m}{m^2 + q^2} \pm \sqrt{ \frac{4 m^2}{(m^2 + q^2)^2}
+ 4 \mu} \right].
\end{equation}
In this case, the dependence $\Gamma^2(q)$ also has a two-peak structure because 
$\Gamma = 0$ at the resonance but $\Gamma^2 > 0$ in its neighborhood. The 
instability is always present for any finite value of $\varepsilon$.

Our explicit solution shows that, if $\alpha > -1$, the instability always occur for 
disturbances with $q$ and $m$ close to the condition of magnetic resonance, 
$\omega_A =0$. The axial field cannot suppress the instability which occurs 
even if $B_z$ is significantly greater than   $B_{\varphi 0}$. 

\subsection{Numerical results}
{
In spite of  the various approximations which have been done, the picture emerging
in the previous session gives a qualitatively correct
account of the MHD spectrum}. In order to show this we 
solved numerically Eq.~(5), assuming  $\alpha=1$ so that $B_{\varphi} \propto r$. 
The results for other profiles of $B_{\varphi}$ are qualitatively similar. 
Eq.~(5) together with the boundary conditions is a two-point boundary value 
problem which can be solved by using the ``shooting'' method \cite{pr92}. In order 
to solve Eq.~(5), we used a fifth-order Runge-Kutta integrator embedded in a 
globally convergent Newton-Rawson iterator. We have checked that the eigenvalue was 
always the fundamental one, as the corresponding eigenfunction had no zero except 
that at the boundaries. 

Fig.~1 exhibits the growth rate of instability as a function of $q$ in the case 
when  the toroidal field is stronger than the axial one ($\varepsilon=0.1$). 
We plot $\Gamma$ for two values of the azimuthal wavenumber, $m=1$ and $m=100$. 
Calculations confirm that only the modes are unstable with the axial wavevectors 
$q$ close to the condition of the magnetic resonance. The resonance values 
of $q= -m/ \varepsilon$ are 10 and 1000 for $m=1$ and $m=100$, respectively. Also, 
in complete agreement with the analytic results (see Eq.~(9)), the growth rate goes 
to 0 at the resonance but $\Gamma^2$ can be positive in its neighborhood. The 
dependence in Fig,~1 is very sharp: the ratio $\delta$ of the half-thickness of 
the peak to $q= -m/ \varepsilon$, corresponding to the resonance, is $\sim 2$ for 
$m=1$ but rapidly decreases and reaches $\approx 0.02$ for $m=100$. The maximum 
growth rate slowly increases with $m$ and becomes $\sim 1$ for large $m$ that 
corresponds to the growth time of the order of the Alfv\'en crossing time. 
\begin{figure}
\begin{center}
\includegraphics[width=9cm]{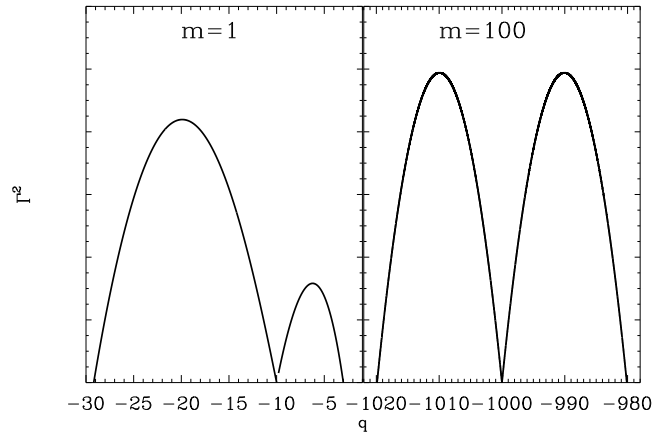}
\caption{The growth rate as a function of $q$ for $\varepsilon=0.1$ and $\alpha=1$.
The panels correspond to $m=1$ and $m=100$. The horizontal axis has different scales
in the panels.}
\end{center}
\end{figure}
In Fig.~2, we plot the dependence of $\Gamma^2$ on $q$ for the same $\alpha$ and 
$\varepsilon=10$. Qualitatively, the behavior of $\Gamma^2$ is similar to that
shown in Fig.~1: only modes with $q$ close to the magnetic resonance can be unstable,
the corresponding range of $q$ is narrow and the instability has a resonance character,
a two-peak structure of $\Gamma^2$ near the resonance, the maximum growth rate increases 
with $m$, etc. Numerically, however, the results differ substantially. The resonance 
peaks are much sharper for $\varepsilon=10$. For example, $\delta$ is $\sim 0.2$\% and 
$\sim 0.1$\% for $m=200$. The maximum growth rate is approximately 
10 times lower than in the previous figure but still is sufficiently high. Note that, 
generally, disturbances with such small wavelengths in the $\varphi$-and $z$-directions 
can be influenced by dissipation (viscosity, 
resistivity). In astrophysical bodies, however, the ordinary and magnetic Reynolds 
numbers are huge and even disturbances with $m \sim 10^2 - 10^4$ can be treated, 
neglecting dissipation. 
\begin{figure}
\includegraphics[width=9cm]{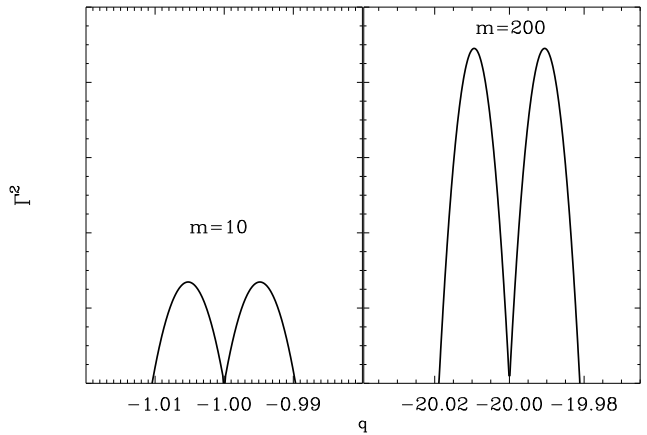}
\caption{Same as in Fig.~1, but for $\varepsilon=10$. The panels correspond to
$m=10$  and $m=200$.}
\label{due}
\end{figure}
\section{Direct numerical simulations}
\begin{figure}
\includegraphics[width=9.cm]{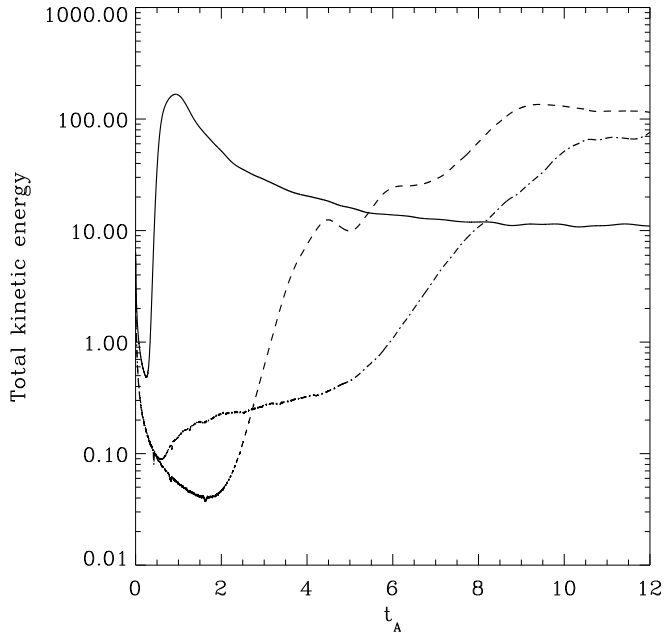}
\caption{Evolution of the mean kinetic density as a function of the Alfv\'en  travel time in the azimuthal directions for all
the three models, $\varepsilon=0$ (solid line), 1 (dashed), and 2 (dot-dashed)}
\label{kene}
\end{figure}
\begin{figure}
\includegraphics[width=8.cm]{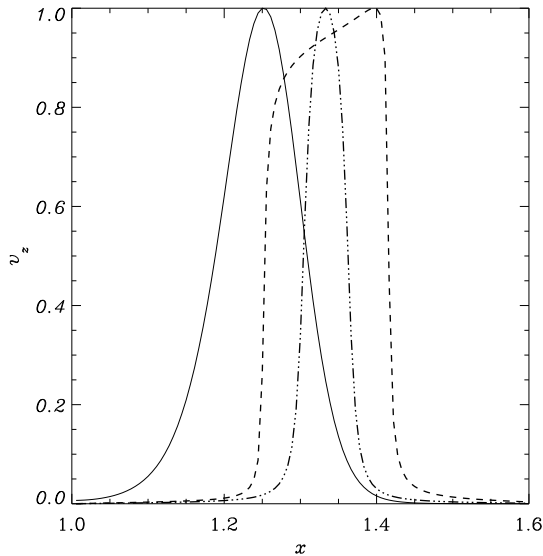}
\caption{Radial profile of the eigenfunctions for the fastest growing modes excited in the simulations from the linear analysis.  
Solid line represent the $\varepsilon=0$ model with $m=1$ and $q\approx 20$. Dashed line and dot-dashed line 
represent the $\varepsilon=1$ case for $m=4$ and $m=6$ respectively at the resonance.}
\label{eigen}
\end{figure}
It is not difficult to realize this type of instability in numerical simulations, at 
least for moderate values of $m$.  
{
In particular, we solved the ideal MHD simulation 
by means of the ZEUSMP code \cite{hayes} in the limit of subthermal field.}
Our setup consists in an isothermal cylinder with a radial extent 
from $s_{\rm in}$ to $s_{\rm out}$ and vertical size $h$ and solve the time dependent 
ideal MHD equations  with periodic boundary conditions in $z$, reflection in $s$ and 
periodic in $\varphi$  and a resolution ranging 
from $120^3$ to $240^3$ all the directions. 
%
The azimuthal field in the basic state is taken in the form
\begin{equation}\label{bs}
B_\varphi= b_0 \; (s/s_0) \exp [-(s-s_0)^2/d^2]
\end{equation}
with $b_0$ being a normalization constant; the axial field $B_z$ is a constant 
whose value can be varied. 
{
In the basic state, the Lorentz force is balanced
with a gradient of pressure, and we have checked that our 
setup was numerically stable if no perturbation was introduced in the system.} 
For actual calculation we have chosen $h=10$, $s_{\rm in}=1.5$, $s_{\rm out}=3$, 
$s_0=2$ and $d^2=0.15$; 
{
the sound speed is assumed to be  much larger than the Alf\'en speed 
($\approx$ ten times), in order to compare our results with the linear 
analysis of the previous session obtained for an incompressible plasma.
After few time steps we perturbed the density with random perturbations in order 
to excite the unstable modes and study their evolution. 
In the case of $\epsilon=0$  the spectrum is dominated by the $m=1$ mode during the linear phase
and we obtain  $\Gamma\approx 11.7$ for the growth rate in units of the Alfv\'en travel time 
in the azimuthal direction. In order to compare this value with the the linear spectrum we explicitly solved Eq.(\ref{pert}) 
for our basic state (\ref{bs}) for various values of $m$ and $q$ obtaining
$\Gamma\approx 13.5$ for the fastest growing modes for the vertical wave 
numbers excited in the numerical simulations according to the spectral analysis.
We found about  $\sim 15\%$ difference with the linear result, we think this discrepancy is acceptable as 3D simulations are usual
rather diffusive and one expects that the actual growth rate should be smaller than the one obtained from
linear analysis.  Similar considerations apply for the $\epsilon>1$ cases. For instance for $\epsilon=1$ 
we find the the fastest growing mode has $\Gamma\approx 1.54$  with  $m=4$ and $m=6$ 
both excited, while the growth rate obtained from the linear analysis predicts $\Gamma\approx 1.45$.
The model with $\varepsilon=2$ has instead $m=9$ as the fastest growing modes and also in this case
the difference with the linear analysis is about $10-15 \%$. The eigenfunctions corresponding 
to the fastest growing modes for $\varepsilon=0$ and $\varepsilon=1$ are depicted in Fig.(\ref{eigen}).
}
In Fig.(\ref{kene}) 
the evolution of the mean kinetic energy 
are plotted as a function of the Alfv\'en travel time.
The solid line is for $\varepsilon=0$, 
while the dashed is for $\varepsilon=1$ and the dot-dashed for 
$\varepsilon=2$. 
Note that  $E_{\rm ax}/E_{\rm tor}\sim 13$ for model $\varepsilon=1$
and $E_{\rm ax}/E_{\rm tor}\sim 42$ for model $\varepsilon=2$ in our setup. The 
growth time for model $\varepsilon=0$ is of the order of the Alfv\'en crossing 
time, while it is significantly longer for models $\varepsilon=1$ and $\varepsilon=2$. 
Nevertheless, the key point that should be 
stressed here is that  the strength of the (turbulent) magnetic energy and turbulent 
kinetic energy at the beginning of the non-linear phase is essentially the same for
all the three models.
Moreover, in the presence of 
a nonzero axial field the corresponding spectrum along the vertical direction 
shows a specific excited mode, so that the resonance condition $q\sim -m/\varepsilon$ 
is satisfied. For model $\varepsilon=2$ for instance, $q\approx 4-5$, for the radial component of 
the magnetic field during the linear evolution.
\begin{figure}
\begin{center}
\hspace{-1cm}\includegraphics[width=9cm]{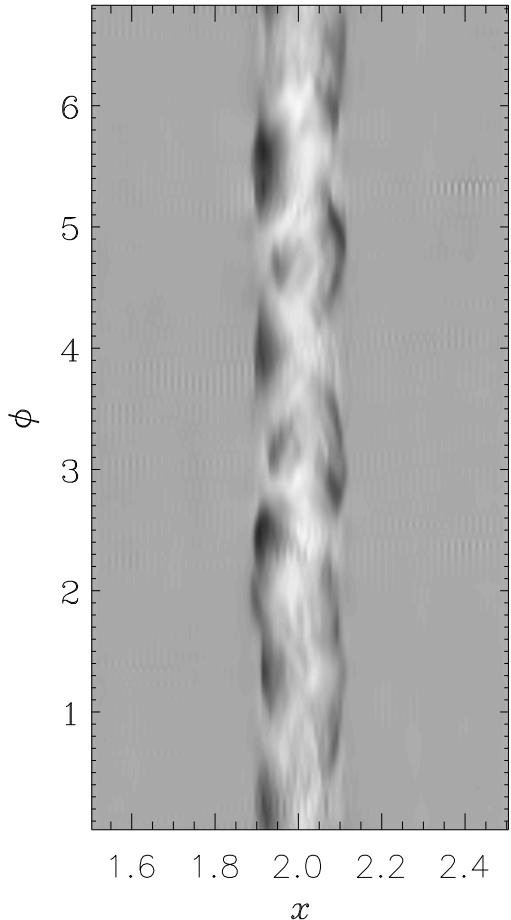}
\end{center}
\caption{The density for model $\varepsilon=2$ during the unstable evolution, around 
$t_A=7.3$, along the plane $z=0$ as a function of radial and azimuthal coordinate. 
The presence of a higher $m$ mode around $m\sim 9$ is clearly visible.  The 
domain along $\varphi$  is $2\pi$ and the resolution of the simulation along the $(z,r,\varphi)$ box is $240 \times 120^2$.}
\label{density}
\end{figure}
Fig.(\ref{density}) shows the occurrence of high $m$ modes for the density in the 
$(s,\varphi)$ plane for $\varepsilon=2$ for a $240\times 120^2$ simulation.
It is difficult to reproduce the instability for much higher values of 
$\varepsilon$. As it is clear from Fig.(\ref{due}) the width of the resonance is 
quite narrow in this case, the growth rate is significantly different from zero 
only for very large values of $m$ and the resolution in all three directions 
needed to reproduce the instability can be extremely large.


\section{Astrophysical implications and conclusions}
{ 
In this paper, we revisited the stability properties of the
screw pinch, a problem which has received considerable attention
in the past in the context of MHD plasma stability for thermonuclear fusion.
As it was  pointed out by Freidberg \cite{frei70}, Eq.(2) describes various 
types of modes which can become unstable under certain conditions.
The basic properties of the unstable modes are similar 
to  those of quasi-kinks and quasi-interchanges obtained by 
\cite{goed71b,goed72} for compressible plasma.
However,  astrophysical condition like those of  stellar interior imply a 
high $\beta$ plasma parameter, a regime which is very far from the 
typical laboratory conditions. To the best of our knowledge, 
an instability of this type has not yet been extensively studied for a
pressure balanced mixed poloidal/toroidal field configuration in the incompressible limit, 
an approximation which can be applied to various astrophysical situations.
The following properties characterize the instability in this case:
i) the instability does not occur for a current-free magnetic configuration;
%
ii) it can arise on a time scale comparable to the
Alfv\'en time scale whereas the growth rate calculated by \cite{goed72}
is an order of magnitude lower, at least; 
iii) the eigenfunctions for high values of $m$ have a resonant character
being very localized as shown in Fig.(\ref{eigen}) for $m=6$;
iv) the dependence of the growth rate on $m$ seems also to be rather peculiar.
In the case of the instability described in \cite{goed72}, unfortunately, the growth rate is 
calculated only in the so called tokamak approximation $B_{\varphi}/xB_z 
\ll 1$ (see Eqs.(30)-(31) by \cite{goed72}) and increases approximately 
proportional to $m$ or even faster. In our case, the dependence on $m$ 
is qualitatively different  because the growth rate saturates 
with $m$ very rapidly, as noticed  in the numerical investigation 
and in the approximate expression (\ref{sette}).

In spite of these differences, quasi-kink and quasi-interchange instabilities 
obtained by \cite{goed71b,goed72} also have the typical double-peak structure 
depicted in Fig.(1) and Fig.(2) as a function of the the axial wavevector.
}

Note that the basic state in our model is characterized by the negative 
pressure gradient in some fraction of the volume, at least. Indeed, hydrostatic 
equilibrium with the toroidal field (10) implies that
\begin{equation}
\frac{d P}{d s} = - \frac{B_{\varphi}^2}{2 \pi s} \left(1 -  
\frac{s(s-s_0)}{d^2} \right)
\end{equation}
Then, $d P/d s < 0$ if $d^2 > s(s-s_0)$. The condition $d P/d s < 0$ is 
required for the development of instability (see, e.g., \cite{long}). 
The sign of the pressure gradient is important because it determines the 
destabilizing effect in the so called Suydam's criterion \cite{suy}. This criterion 
represents a necessary but local condition for stability and it reads in our notations 
\begin{equation}
\frac{s B_z^2}{4 \pi} \left( \frac{1}{h} \frac{d h}{d s} \right)^2 
+ 8 \frac{d P}{d s} > 0 ,
\label{12}
\end{equation}
where $h = s B_z / B_{\varphi}$ is the magnetic shear. In the case of the
basic state with toroidal field (10), the necessary condition for stability
is not satisfied in some fraction of the volume (for example, in a neighborhood 
of $s_0$). This violation of the stability condition (\ref{12}) is actually indicating
the presence of at least some unstable mode in the system. 

{
Stability properties of magnetic configurations are of great importance for 
various astrophysical applications. For instance, it is widely believed that 
magnetic fields play an important role in the formation and propagation of 
astrophysical jets providing an efficient mechanism of collimation through 
magnetic tension forces (e.g., \cite{blan}). Polarization observations provide 
information on the orientation and degree of order of the magnetic field in jets.
It appears that many jets can develop relatively highly organized magnetic 
structures. To explain the observational data, various simplified models of  
three-dimensional magnetic structures have been proposed. Typically, the magnetic 
field can have both longitudinal component and substantial toroidal component in 
the core region (see, e.g., \cite{gab}). The mechanisms responsible for 
generation of the magnetic field in jets are still unclear. Since the origin 
of jets is probably relevant to MHD-processes in magnetized plasma, their 
magnetic fields could be generated during the process of jet formation (see, 
e.g., \cite{rom}) or, alternatively, it can be generated by the dynamo 
mechanism \cite{urp} when the jet propagates in the interstellar medium. 
In both cases, the stability is a crucial issue for the properties of the jet.
For instance, the origin of relatively small scale structures within the jet 
can be attributed to different instabilities arising in jets, including the one
considered in our study. Magnetic structures that appears as a result of the
development of instabilities can manifest themselves in polarization observations
of the jets. 

The considered instability can play an important role in magnetic stars where
it can affect the magnetic field in stably stratified regions. Spruit \cite{spr} 
reviewed various types of instabilities that are likely to intervene in a 
magnetized radiative regions of stars, and he concluded that the strongest 
among them are those which are related to the instability of magnetic 
configurations. According to \cite{spr}, turbulence generated by such 
instability can drive a genuine dynamo in stellar radiative zones (see, 
however, \cite{zahn}). Understanding the conditions required for the instability
is, therefore, crucial for dynamo models in stably stratified zones of
stars. 

This type of magnetic instabilities can be of interest also for neutron stars where 
the magnetic field reaches an extremely high value $\sim 10^{13}-10^{14}$ G. 
Such a strong field can be generated by the turbulent dynamo action during 
the very early stage of evolution (see \cite{bon}) when the neutron star is 
convectively unstable. This unstable stage lasts less than $\sim 1$ min. 
The further evolution of the magnetic field is determined mainly by ohmic 
dissipation but can be affected by current-driven instabilities as well
\cite{land} because dynamo in the convective zone generates a magnetic 
configuration that is not equilibrium.   
}

\noindent
{\it Acknowledgments.}
VU thanks INAF-Ossevatorio Astrofisico di Catania for hospitality and
financial support. All the computations were performed on the sanssouci-cluster
of AIP whose support is gratefully acknowledged.


\vspace{-0.5cm}

\begin{thebibliography}{}

\vspace{-0.5cm}
 
 
\bibitem{frei}
Freidberg, J.P.: Ideal Magnetohydrodynamics. Plenum Press (1987); 

\bibitem{bernstein} 
Bernstein, I.B., Frieman, E.A., Kruskal, M.D., Kulsrud, R.M., Proc. R. Soc. A244, 17 (1958)
 
\bibitem{wr73}
Wright G.A.E. Mon. Not. R. astr. Soc. 162, 339 (1973)

\bibitem{tay73a}
Tayler R.J. Mon. Not. R. astr. Soc. 161, 365 (1973)

\bibitem{tay73b}
Tayler R.J. Mon. Not. R. astr. Soc. 163, 77 (1973)

\bibitem{tay80}
Tayler R.J. Mon. Not. R. astr. Soc. 191, 151 (1980)

%
\bibitem{gopo}
Goedbloed, H., Poedts, S., Principles of Magnetohydrodynamics, CUP, (2004) 

\bibitem{deblank}
de Blank H.J., Transaction of Fusion science and Technology vol. 4 feb. 2006,

\bibitem{bo08a}
Bonanno A., Urpin V. Astron. Astrophys. 477, 35 (2008)

\bibitem{bo08b}
Bonanno A., Urpin V. Astron. Astrophys. 488, 1 (2008)

\bibitem{bs04}
Braithwaite J., Spruit H. Nature 431, 819 (2004)

\bibitem{bn06}
Braithwaite J., Nordlund A. Astron. Astrophys. 450, 1077 (2006)

\bibitem{br08}
Braithwaite J. Mon. Not. R. astr. Soc. 386, 1947 (2008)

\bibitem{br09}
Braithwaite J. Mon. Not. R. astr. Soc. 397, 763 (2009)

\bibitem{duez}
Duez, V.; Braithwaite, J.; Mathis, S., ApJ 724L, 24 (2010)

\bibitem{bo11}
Bonanno A., Urpin V. Astron. Astrophys. 525, 100 (2011)

\bibitem{frei70}
Freidberg J. Phys. Fluids. 13, 1812 (1970)

\bibitem{goed71}
Goedbloed J.P. Physica. 53, 501 (1971)

\bibitem{pr92}
W.H.Press, S.A.Teukolsky, W.T.Vetterling, and B.P.Flannery. {\it Numerical 
Recipes in FORTRAN. The art of scientific computing} (Cambridge UP, 1992). 

\bibitem{hayes}
Hayes, J. C., Norman, M. L., Fiedler, R. A., Bordner, J. O., Li, P. S., et al.  ApJS, 165, 188 (2006)

\bibitem{goed71b}
Goedbloed J.P. Physica. 53. 535 (1971)

\bibitem{goed72}
Goedbloed J.P., Hagebeuk H. Phys. Fluids. 15, 1090 (1972)

\bibitem{long}
Longaretti P.-Y. 2003. PhLA, 320, 215

\bibitem{suy}
B.R. Suydam, in: Proc. of the Second U.N. Internat. Conf. on the Peaceful Uses of Atomic Energy, Vol. 31, United
Nations, Geneva, 1958, p. 157.

\bibitem{blan}
Blandford R. 1993. In "Astrophysical Jets" (Eds. D.Burgarella,
M.Livio \& C.P.O'Dea), Cambridge: Cambridge University Press 
K\"{o}nigl A., Pudritz R. 1999. In "Protostars and Planets III" (Eds.
V.Mannings, A.Boss \& S.Russell), Tucson: University of Arizona
Press

\bibitem{gab}
Hirabayashi H. et al. 1998. Science, 281, 1825;
Gabuzda D., Murray E., Cronin P. 2004. MNRAS, 351, 89L

\bibitem{rom}
Blandford R., Payne D. 1982. MNRAS, 199, 883; 
Romanova M., Lovelace R. 1992. A\&A, 262, 26;
Koide S., Shibata K., Kudoh T. 1998. ApJ, 495, L63

\bibitem{urp}
Urpin V. 2006. A\&A, 455, 779

\bibitem{spr}
Spruit H. 1999. A\&A, 349, 189

\bibitem{zahn}
Zahn J.-P., Brun A., Mathis S. 2007. A\&A, 474, 145

\bibitem{bon}
Bonanno A., Rezzolla L., Urpin V. 2003, A\&A, 410, 33; 
Bonanno A., Urpin V., Belvedere G. 2005. A\&A, 440, 199;
Bonanno A., Urpin V., Belvedere G. 2006. A\&A, 451, 1049

\bibitem{land}
Lander S.K., Jones D.I. 2011. MNRAS, 412, 1394; Kiuchi K., 
Yoshida S., Shibata M. 2011. astro-ph/1104.5561


%
%
%
%
%
%

\end{thebibliography}
\end{document}